\documentclass[conference]{IEEEtran}
\usepackage{etoolbox}
\makeatletter
\patchcmd{\@makecaption}
  {\scshape}
  {}
  {}
  {}
\makeatother
\IEEEoverridecommandlockouts
\usepackage{cite}
\usepackage{amsmath,amssymb,amsfonts}
\newtheorem{assumption}{Assumption}[section]
\newcommand\independent{\protect\mathpalette{\protect\independenT}{\perp}}
\def\independenT#1#2{\mathrel{\rlap{$#1#2$}\mkern2mu{#1#2}}}
\usepackage{graphicx}
\usepackage{subfigure}
\usepackage{textcomp}
\usepackage{xcolor}
\usepackage{xspace}
\usepackage{multirow}
\usepackage{enumitem}
\usepackage{algorithm} 
\usepackage{algpseudocode} 
\usepackage{mathtools}
\usepackage{hyperref}
\def\BibTeX{{\rm B\kern-.05em{\sc i\kern-.025em b}\kern-.08em
    T\kern-.1667em\lower.7ex\hbox{E}\kern-.125emX}}

\newcommand{\model}{\texttt{DSW}\xspace}
\newcommand{\modelnospace}{\texttt{DSW}}

\begin{document}

\title{Estimating Individual Treatment Effects with Time-Varying Confounders\\
}

\author{\IEEEauthorblockN{Ruoqi Liu\IEEEauthorrefmark{1},
Changchang Yin\IEEEauthorrefmark{1},
Ping Zhang\IEEEauthorrefmark{1}
}
\IEEEauthorblockA{\IEEEauthorrefmark{1}The Ohio State University, Columbus, Ohio, USA 43210\\ Email:
\{liu.7324, yin.731, zhang.10631\}@osu.edu}
}


\maketitle

\begin{abstract}
Estimating the individual treatment effect (ITE) from observational data is meaningful and practical in healthcare. Existing work mainly relies on the \textit{strong ignorability assumption} that no hidden confounders exist, which may lead to bias in estimating causal effects. Some studies considering the hidden confounders are designed for static environment and not easily adaptable to a dynamic setting. In fact, most observational data (e.g., electronic medical records) is naturally dynamic and consists of sequential information. In this paper, we propose Deep Sequential Weighting (\modelnospace) for estimating ITE with time-varying confounders. Specifically, \model infers the hidden confounders by incorporating the current treatment assignments and historical information using a deep recurrent weighting neural network. The learned representations of hidden confounders combined with current observed data are leveraged for potential outcome and treatment predictions. We compute the time-varying inverse probabilities of treatment for re-weighting the population. We conduct comprehensive comparison experiments on fully-synthetic, semi-synthetic and real-world datasets to evaluate the performance of our model and baselines. Results demonstrate that our model can generate unbiased and accurate treatment effect by conditioning both time-varying observed and hidden confounders, paving the way for personalized medicine.



\end{abstract}

\begin{IEEEkeywords}
deep learning, electronic medical record, ITE, time-varying confounders
\end{IEEEkeywords}
\section{Introduction}
Estimating the individual treatment effect (ITE) is a task of evaluating the causal effect of treatment strategies on some important outcomes over individual-level, which is a significant problem in many areas \cite{glass2013causal,baum2015causal,wang2015robust,heckman2018returns,song2019effect}. For example, in healthcare domain, it is critical to prescribe personalized medicines (treatments) for different patients based on their health conditions. 
To approach this, randomized controlled trial (RCT) is usually conducted, which is accomplished by randomly allocating patients to two groups, treating them differently (i.e., one group has intervention and the other has a placebo or no intervention) and comparing them in terms of a measured response. However, conducting RCTs in the healthcare domain is extremely expensive and time-consuming, if not impossible, due to the requirement of tremendous expert effort and the consideration of ethical issues.

Observational data contain patient records, including their demographic information, vital signs, lab tests and outcomes but without having complete knowledge of why a specific treatment is applied to a patient.
The accumulation of observational data in electronic medical records (EMRs) offers a promising opportunity for ITE estimation when RCTs are expensive or impossible to conduct.

\begin{figure}[t]
\centering
\includegraphics[width=0.9\linewidth]{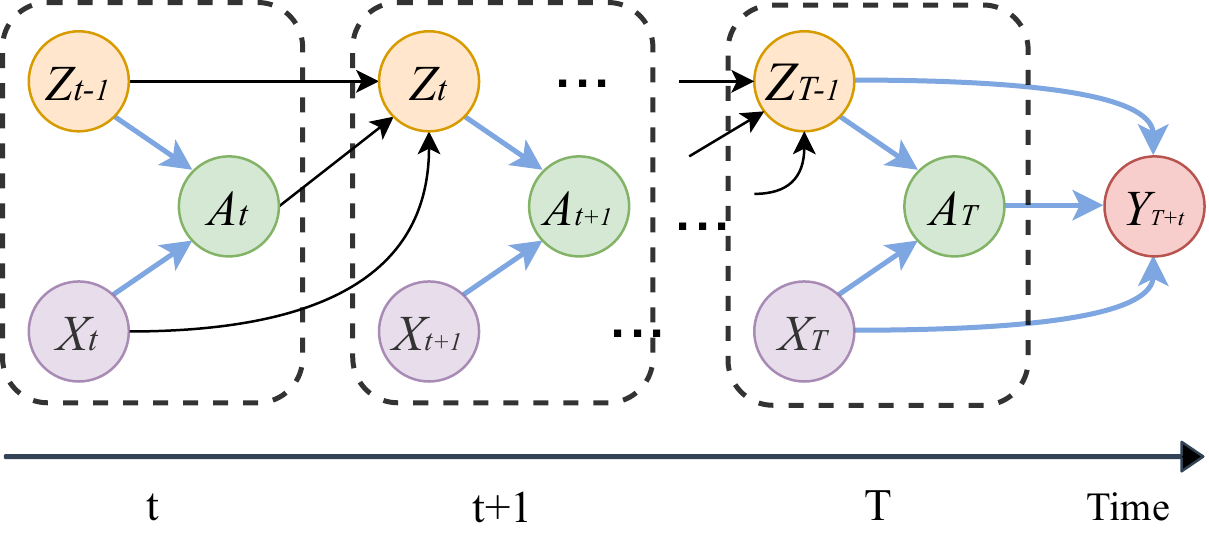}
\caption[]{The illustration of causal graphs for causal estimation from dynamic observational data. During observed window $T$, we have a trajectory $\{X_t, A_t, Z_{t-1}\}_{t=1}^{T}\cup\{Y_{T+\tau}\}$, where $X_t$ denotes the current observed covariates, $A_t$ denotes the treatment assignments, $Z_t$ denotes the hidden confounders and $Y_{T+\tau}$ denotes the potential outcomes. At any time stamp $t$, the treatment assignments $A_t$ are affected by both observed covariates $X_t$ and hidden confounders $Z_{t-1}$ (the causal relationship are annotated by blue arrows). The hidden confounders $Z_{t-1}$ are inferred from previous hidden confounders, treatments and covariates (annotated by black arrows). The potential outcomes $Y_{T+\tau}$ are affected by last time stamp covariates $X_{T}$, treatment assignments $A_{T}$ and hidden confounders $Z_{T-1}$.}
\vspace{-15pt}
\label{problem_setting}
\end{figure}

There have been lots of existing methods that estimate ITE by leveraging observational data, including propensity score matching method \cite{rosenbaum1983central}, forest based methods \cite{hill2011bayesian,wager2018estimation}, and representation learning-based methods \cite{johansson2016learning,shalit2017estimating,yao2018representation}. However, many methods are mainly based on the \textit{strong ignorability assumption} that there are no unobserved confounders and only few studies have considered the influence of hidden confounders \cite{louizos2017causal}. Here, the hidden confounders refer to factors that affect both treatment assignment and outcome, but are not directly measured in the observational data. For example, physicians may prescribe treatments to the patient based on indicators not in the medical records. Ignoring these hidden confounders can lead to bias in estimating causal effects \cite{pearl2009causality}. 
Besides, these methods are primarily designed for static settings and are hardly extended to longitudinal observation data. However, most real-world observational data is naturally dynamic and consists of sequential information. For example, in EMR, the patient's conditions (e.g., prescribed medicines, lab results and vital signs) and treatment assignments are recorded frequently during their stay in the hospital. Therefore, estimating ITE becomes more challenging when the treatments and covariates change over time, and the potential outcomes are influenced by historical treatments and covariates.


In this paper, we study the problem of \textit{Estimating individual treatment effects with time-varying confounders} (as illustrated by a causal graph in Fig \ref{problem_setting}). To alleviate the aforementioned challenges, we propose a novel causal effects estimation framework, Deep Sequential Weighting (\modelnospace), to adjust the time-varying confounders in the longitudinal data. The proposed framework \model consists of three main components: representation learning module, balancing module and prediction module. To adjust the time-varying hidden confounders, \model first learns the representations of hidden confounders by leveraging the current observed covariates and all historical information (i.e., previous covariates and treatment assignments) through Gated Recurrent Units (GRU) with an attention mechanism. With the help of the attention mechanism, the model can automatically focus on important historical information. Then, we compute the time-varying inverse probability of treatment for each individual to balance the confounding. The learned representations of hidden confounders and observed covariates are then combined together for both treatment prediction and potential outcome prediction.

To demonstrate the effectiveness of our framework, we conduct comprehensive experiments on synthetic, semi-synthetic and real-world EMR datasets (MIMIC-III \cite{johnson2016mimic}). \model outperforms state-of-the-art baselines in terms of PEHE and ATE.
To further illustrate how our method can be used in personalized medicine,  we analyze the treatment effects on important outcomes for ICU septic patients. Results demonstrate that our model can generate unbiased and accurate treatment effect by conditioning both time-varying observed confounders and hidden confounders. Our model has the potential to be leveraged as part of clinical decision support systems that assist physicians to determine whether to introduce treatment to a patient or a specific population. 

The contributions of this paper are as follows:
\begin{itemize}
  \item We study the task of estimating ITE with time-varying confounders, on which few attention has been before. 
  \item We propose a novel causal inference framework \model to solve the task. \model fully utilize the historical information and current covariates for learning the representations of hidden confounders. A balancing operation is adopted to generate unbiased and accurate ITE estimation.
  \item We conduct experiments on synthetic, semi-synthetic and real-world datasets to demonstrate the effectiveness of our proposed method. Results show that our method outperforms state-of-the-art causal inference methods and 
  has the potential to be used as part of clinical decision support systems to determine whether a treatment is needed for a specific patient, paving the way for personalized medicine.
\end{itemize}

\section{Method}
In this section, we first give a formal definition of the notations used throughout the paper, then present the proposed framework for estimating ITE.
\subsection{Preliminary}
Let $X_t\in\mathcal{X}_t$ be the time-dependent covariates of the observational data at time stamp $t$ such that $X_t=\{x^{(1)}_{t}, x^{(2)}_{t},...,x^{(n)}_{t}\}$, where $x^{(i)}_{t}$ denotes the covariates for $i$-th patient, $n$ denotes the number of patients, and $\mathcal{X}_t$ denotes the time-dependent feature space. The static features (e.g., demographic information), do not change overtime are also considered as observed covariates. We use $C\in\mathcal{C}$ represent the static features for all the patients. At each time stamp $t$, the treatment assignments are denoted as $A_{t}=\{a^{(1)}_{t},a^{(2)}_{t},...,a^{(n)}_{t}\}$, where $A_{t}\in\mathcal{A}$, $a^{(i)}_{t}$ denotes the treatments assigned to $i$-th patient. In the case of the binary treatment setting, i.e., $a^{i}_{t}=\{0, 1\}$, where $1$ is considered as "treated" while $0$ as "control", we are interested in estimating the effect of the treatment assigned until time stamp $T$ on the outcomes $Y_{T+\tau}\in\mathcal{Y}$, observed at time stamp $T+\tau$, where $\tau$ is the prediction window. Note that in observational data, a patient can only belong to one group (i.e., either treated or control group), thus the outcome from the other group is always missing and referred to counterfactual. To represent the historical sequential data before time stamp $t$, we use the notation $\overline{X}_{t}=\{X_{1},X_{2},...,X_{t-1}\}$ to denote the history of covariates observed before time stamp $t$, and $\overline{A}_{t}$ refers to the history of treatment assignments. Combining all covariates and treatments, we define $\tilde{H}^{(i)}_{t}=\{\overline{x}^{(i)}_{t},\overline{a}^{(i)}_{t}\}\cup\{c^{(i)}\}$ as all the historical data collected before time stamp $t$. The observational data for $i$-th patient can be represented using the notations defined above as: $\mathcal{D}^{(i)}=\{x^{(i)}_{t},a^{(i)}_{t}\}^{T}_{t=1}\cup\{c^{(i)},y^{(i)}_{a,T+\tau}\}$. We summarize the notations we used in this paper in Table \ref{tab:notations}.

\begin{table}[t]
\caption{Notations}
\label{tab:notations}
    \centering
    {\renewcommand{\arraystretch}{1.1}%
    \begin{tabular}{ll}\hline
    Notation    &   Definition  \\\hline
    $\mathcal{X}$     &   The space of time-varying covariates \\
    $\mathcal{C}$   &   The space of static covariates   \\
    $\mathcal{A}$   &   The set of treatment options of interest   \\
    $\mathcal{Y}$   &   The space of potential outcomes  \\
    $X_{t}$ &   The time-varying covariates of all patients at time $t$  \\
    $\mathcal{Z}$   &   The space of hidden confounders \\
    $x^{(i)}_{t}$   &   The time-varying covariates of $i$-th patient at time $t$  \\
    $C$   &   The static covariates of all patients \\
    $c^{(i)}$ &   The static covariates of $i$-th patient \\
    $A_{t}$ &   The treatment assigned at time $t$  \\
    $a^{(i)}_{t}$   &   The treatment assigned for $i$-th patient at time $t$  \\
    $Y_{T+\tau}$    &   The factual (observed) outcomes at time $T+\tau$ \\
    $Y_{1,T+\tau}/\hat{Y}_{1,T+\tau}$   &   
    \begin{tabular}[c]{@{}l@{}}The observed/predicted outcome at time $T+\tau$ when \\ receive treatment\end{tabular}\\
    $y^{(i)}_{1,T+\tau}/\hat{y}^{(i)}_{1,T+\tau}$   &   
    \begin{tabular}[c]{@{}l@{}}The observed/predicted outcomes of $i$-th patient at \\ time $T+\tau$ when $a^{(i)}=1$\end{tabular}\\
    $Y_{0,T+\tau}/\hat{Y}_{0,T+\tau}$   &  
    \begin{tabular}[c]{@{}l@{}} The observed/predicted outcome at time $T+\tau$ when \\ not receive the treatment\end{tabular}\\
    $y^{(i)}_{0,T+\tau}/\hat{y}^{(i)}_{0,T+\tau}$   &   
    \begin{tabular}[c]{@{}l@{}} The observed/predicted outcomes of $i$-th patient at \\ time $T+\tau$ when $a^{(i)}=0$\end{tabular}\\
    $e^{(i)}/\hat{e}^{(i)}$ &   The true/predicted ITE of $i$-th patient at time $t$  \\
    $\overline{(\cdot)}_{t}$ &   The historical covariates collected before time $t$   \\ 
    $Z_t$   &   The learned hidden confounders at time $t$  \\
    $z^{(i)}_{t}$   &   The learned hidden confounders for patient $i$ at time $t$ \\
    $\tilde{H}^{(i)}_{t}$   &      
    \begin{tabular}[c]{@{}l@{}} The historical data consists of
    $\{\overline{x}^{(i)}_{t},\overline{a}^{(i)}_{t}\}\cup\{c^{(i)}\}$ \\ for $i$-th patient\end{tabular}\\
    $T$ &   The number of time stamps (observation window)    \\
    $\tau$  &   The length of prediction window \\
    $n$ &   The number of patients in the dataset\\
    \hline
    \end{tabular}}
\vspace{-10pt}
\end{table}

We follow the well-adopted potential outcome framework and its variation that considers the time-varying treatment assignments when estimating the causal effect on the outcomes. The potential outcome $y^{(i)}_{a,T+\tau}$ of $i$-th patient given the historical treatment can be formulated as $y^{(i)}_{a,T+\tau}=\mathbb{E}[y|x^{(i)}_{t},\mathcal{H}^{(i)}_t,a=a^{(i)}]$, where $a^{(i)}$ equals to 1 if the treatment is assigned at time $\{1,2..,T\}$, otherwise 0. Then the individual treatment effect (ITE) on the temporal observational data is defined as follows:
\begin{small}
\begin{equation}
    e^{(i)}=\mathbb{E}[y^{(i)}_{1,T+\tau}|x^{(i)}_{t},a^{(i)}_{t},\tilde{H}^{(i)}_{t}]-\mathbb{E}[y^{(i)}_{0,T+\tau}|x^{(i)}_{t},a^{(i)}_{t},\tilde{H}^{(i)}_{t}]
\end{equation}
\end{small}
Here, the observed outcome $y^{(i)}_{a,T+\tau}$ under treatment $a$ is called factual outcome, while the unobserved one $y^{(i)}_{1-a,T+\tau}$ is the counterfactual outcome. In observational data, only the factual outcomes are available, while the counterfactual outcomes can never been observed. 

\vspace{-5pt}
\subsection{Assumptions}
Our estimation of ITE is based on the the following important assumptions \cite{hernan2010causal}, and we further extend the assumptions in our scenario (i.e., time-varying observational data). 
\begin{assumption}[Consistency]\label{as:consistency}
The potential outcome under treatment history $\overline{A}$ equals to the observed outcome if the actual treatments history is $\overline{A}$.
\end{assumption}

\begin{assumption}[Positivity]\label{as:positivity}
For any patient $i$, if the the probability $\mathbb{P}(\overline{a}^{(i)}_{t-1},\overline{x}^{(i)}_{t}, c^{(i)})\neq0$, then the probability of receiving treatment $0$ or $1$ is positive, i.e., $0<\mathbb{P}(\overline{a}^{(i)}_{t},\overline{x}^{(i)}_{t}, c^{(i)})<1$, for all $\overline{a}^{(i)}_{t}$.   
\end{assumption}
Besides these two assumptions, many existing work are based on \textit{strong ignorability} assumption:
\begin{assumption}[Strong Ignorability]\label{as:ignorability}
Given the observed historical covariates $\overline{x}^{(i)}_t$ and static covariates $c^{(i)}$ of $i$-th patient, the potential outcome variables $y^{(i)}_{1,T+\tau}$, $y^{(i)}_{0,T+\tau}$ are independent of the treatment assignment, i.e., $(y^{(i)}_{1,T+\tau},y^{(i)}_{0,T+\tau})\independent a^{(i)}_{t}|\overline{x}^{(i)}_t,c^{(i)}$
\end{assumption}

This assumption holds only if there exist no hidden confounders. However, this condition is hard to guarantee in practice especially in real-world observational data. In this paper, we relax such strict assumption by introducing that there exist potential hidden confounders. Our proposed method can learn the representations of the hidden confounders and eliminate the bias between the treatment assignments and outcomes at each time stamp. The learned representations (denoted by $Z_{t}\in\mathcal{Z}$) can be leveraged for inferring the unobserved confounders and regarded as the substitutes of hidden confounders. Thus, we extend the \textit{strong ignorability} assumption by considering the existing of hidden confounders $Z_{t}$ at each time stamp $t$, which influence the treatment assignment $A_t$ and potential outcomes $Y_{T+\tau}$. Given the hidden confounders $Z_{t}$, the potential outcome variables are independent of the treatment assignment at each time stamp. 

\vspace{-5pt}
\subsection{Proposed Method}
\begin{figure*}[t]
\centering
\includegraphics[width=0.8\textwidth]{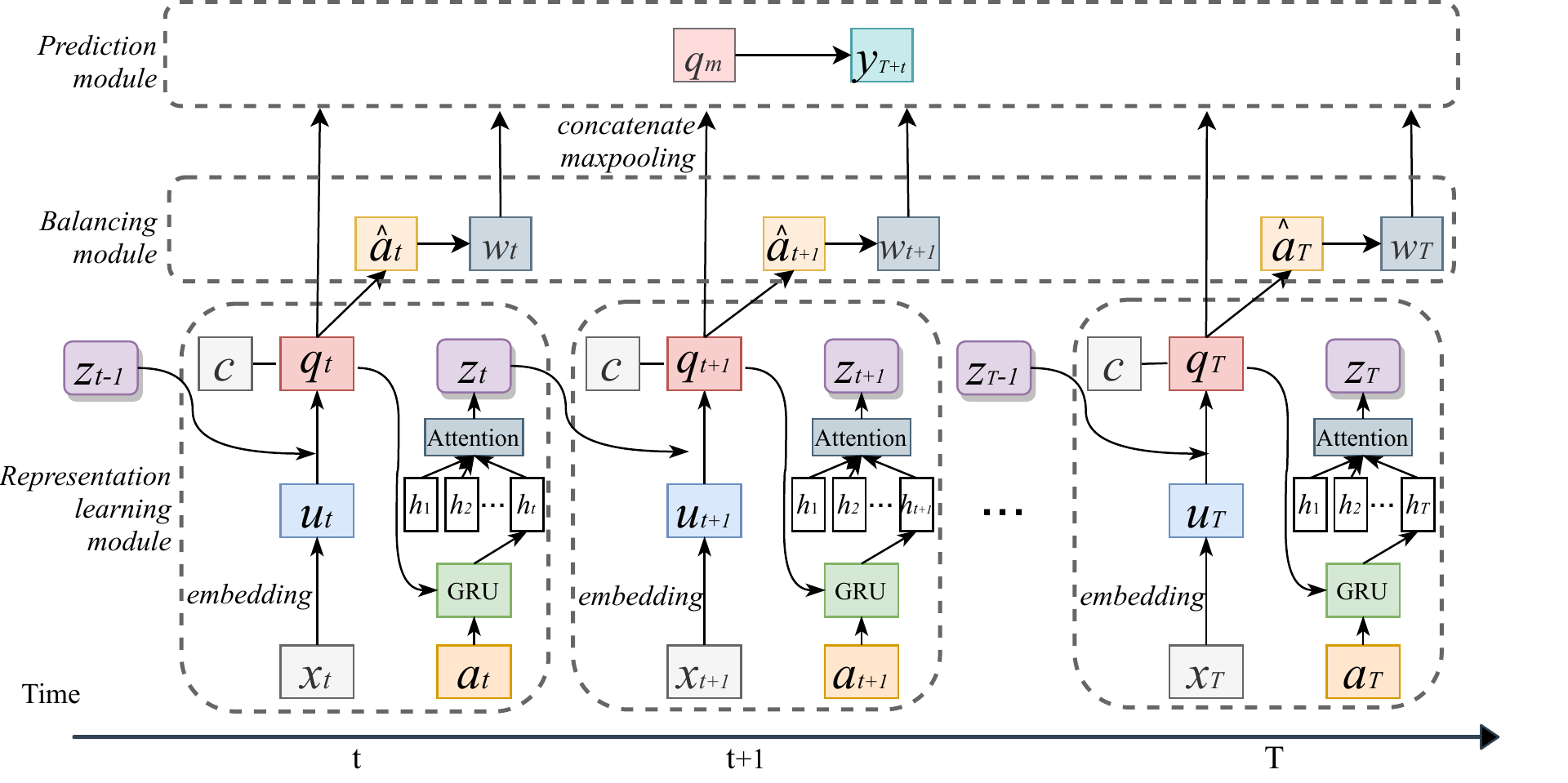}
\vspace{-5pt}
\caption[]{The framework of \model. \model contains three main modules: representation learning module, balancing module and prediction module. At each time stamp $t$, the model takes the current covariates ($X_t$, $C$), treatment assignments $A_t$, along with the hidden variables $Z_{t-1}$ as input for learning representations of confounder $Q_t$. The historical information is modeled via a gated recurrent unit (GRU) and aggregated through an attention layer. Then, we use the learned representations of confounding variables for treatment prediction at each time stamp and final outcome prediction after time $T$.}
\vspace{-10pt}
\label{model}
\end{figure*}
According to the aforementioned assumptions, we present a novel method that utilizes current variables as well as the historical data to learn the representations of the hidden confounders and estimate the individual treatment effect (ITE) for all patients. The overall framework of the proposed method is illustrated in Fig. \ref{model}. We will illustrate the details of each module in the following subsections.

\subsection{Representation Learning Module}
As the initial feature vectors are always high-dimensional and sparse in the case of real-world data, we first convert the initial features $x^{(i)}_t\in\mathbb{R}^{d_x}$ of each patient into a lower-dimensional and continuous data representations $u^{(i)}_t\in\mathbb{R}^{d_u}$, where $d_u$ is the dimension of the embedded feature vectors, using a liner embedding layer. That is, we define:
\begin{equation}
\label{eq:embedding}
    u^{(i)}_t=W_{emb}x^{(i)}_t
\end{equation}
where $W_{emb}\in\mathbb{R}^{d_{u}\times d_{x}}$ is the embedding matrix. Simply, the liner embedding layer can be alternatively replaced by other more complex embedding method such as multi-layer perceptron (MLP) \cite{pal1992multilayer}, which is also widely used in learning representation of EMR data \cite{choi2016multi}. 

The representation of hidden confounders $z^{(i)}_t$ is learned through a GRU layer to capture the inherent characteristics of time-varying observational data (i.e., dependency between each time unit and sparsity). The current information $u^{(i)}_t$, treatment assignments $a^{(i)}_t$, last time stamp hidden confounders $z^{(i)}_{t-1}$ and last output hidden state of GRU $h^{(i)}_{t-1}$ are regarded as the input to the GRU. For the convenience of future prediction task, we concatenate the current information $u^{(i)}_t$ and last time stamp hidden confounders $z^{(i)}_{t-1}$ into a new variable $q^{(i)}_t$ as follows,
\begin{equation}
\label{eq:hidden-representation-confounding}
    q^{(i)}_t=g([u^{(i)}_t,z^{(i)}_{t-1}])
\end{equation}
where $g(\cdot)$ denotes the function for learning the hidden confounders (typically a MLP layer is used for generating the representations), $[\cdot,\cdot]$ denotes the concatenation of two vectors.

We elaborate the the architecture of GRU as follows,
\begin{equation}
\label{eq:gru}
\begin{aligned}
    f_t &=\sigma_{g}(W_{f}[q^{(i)}_t,c^{(i)}_t,a^{(i)}_t]+V_{f}h_t+b_{f})\\
    r_{t} &=\sigma_{g}(W_{r}[q^{(i)}_t,c^{(i)}_t,a^{(i)}_t]+V_{r}h_t+b_{r})\\
    h^{\prime}_t &=\Phi_{h}(W_{h}[q^{(i)}_t,c^{(i)}_t,a^{(i)}_t]+V_{f}(r_{t}\odot h_{t-1})+b_{h})\\
    h_t &=f_t\odot h_{t-1}+(\textbf{1}-f_t)\odot h^{\prime}_t
\end{aligned}
\end{equation}
where $\sigma_{g}$ is sigmoid function, $\Phi_{h}$ is hyperbolic tangent function, $\odot$ denotes element-wise product operation, $W_{f}$, $W_{r}$, $W_{h}\in\mathbb{R}^{d_{h}\times (d_{q}+d_{c}+1)}$, $V_{f}$, $V_{r}\in\mathbb{R}^{d_{h}\times d_{h}}$, $b_{f}$, $b_{r}\in\mathbb{R}^{d_h}$ are parameters matrices and vectors to learn. $f_t$ denotes the update gate vector, $r_t$ denotes reset gate vector and $h_t$ denotes the hidden output vector. The output vectors are further aggregated via a attention layer for automatically focusing on important historical time stamp. We can use various methods to calculate the attention energies between the each previous hidden state $h_{s}$ and current state $h_t$, e.g., dot product $h^{\top}_t h_{s}$, linear attention $h^{\top}_t W_{\alpha} h_{s}$. In this paper, we calculate the attention weight $\alpha_{t,s}$ using a method that concatenates each previous hidden state with the current state, and the product of two states. That is,
\begin{equation}
\label{eq:attention}
\begin{aligned}
    \alpha_{t,s} &=\text{score}(h_t, h_s)=\Phi(W_{\alpha}[h_t,h_s,h_t\odot h_s])\\
    \mathbf{\alpha}_t &=\text{softmax}(\alpha_{t,1},\alpha_{t,2},\dots,\alpha_{t,t-1})
\end{aligned}
\end{equation}
where $\Phi$ is hyperbolic tangent function, $W_\alpha\in\mathbb{R}^{3d_h\times d_h}$ is learnable parameter matrix. Using the generated attention energies, we can calculate the context vector $o_t$ for each patient up to $t$ time stamp as follows,
\begin{equation}
\label{eq:context-vector}
    o_t=\sum_{s=1}^{t-1}\alpha_{t,s}h_s
\end{equation}
We further concatenate the context vector with the current hidden state to generate the current representations $z_t$ up to time $t$:
\begin{equation}
\label{eq:hidden-confounding}
    z_t=\Phi(W_{z}[h_t,o_t])
\end{equation}
where $W_z\in\mathbb{R}^{2d_{h}}$ is a learnable parameter matrix.


\subsection{Prediction Module}
After obtaining the hidden confounding representations, we leverage them for treatment prediction at each time stamp and final outcome prediction during the following prediction window.

\subsubsection{Global Max Pooling}
As the time sequence increases, basic RNN model may forget earlier information due to its long-term dependency. Thus, we adopt a global max-pooling operation over the concatenation of all outputs of $q^{(i)}_t$ vectors. As shown in Fig. \ref{model}, the output of max-pooling layer $q^{(i)}_m$ are further used for treatment prediction and potential outcome prediction.

\subsubsection{Treatment prediction}
We predict the treatment assignments for the patient at each time stamp regarding the $q^{(i)}_m$ and static demographic features $c^{(i)}$ as input, and the real treatment $a^{(i)}_t$ as the target label. The predicted treatments $\hat{a}^{(i)}_t$ are obtained through a fully-connected layer with sigmoid function as the last layer,
\begin{equation}
\label{eq:pred_a}
\hat{a}_t = \text{sigmoid}(W_{a}[q_{m},c]+b_{a})
\end{equation}
where $W_{a}\in\mathbb{R}^{d_{q}+d_{c}}$ and $b_{a}\in\mathbb{R}$ are learnable parameters, $\hat{a}^{(i)}_t$ denotes the the probability of receiving treatment based on the potential confounders of patient $i$ at time $t$. Typically, the predicted results can also be referred as propensity score \cite{rosenbaum1983central} that $\hat{a}^{(i)}_t=\mathbb{P}(a^{(i)}_{t}=1|u^{(i)}_{t},z^{(i)}_{t-1})$.

As we consider the binary treatment in this paper, we use a cross-entropy loss for the treatment prediction over all patients and all time stamps as follows,
\begin{small}
\begin{equation}
\label{eq:treatment-pred-loss}
    \mathcal{L}_a=-\frac{1}{N}\frac{1}{T}\sum_{i=1}^{N}\sum_{t=1}^{T}(a^{(i)}_t\log{\hat{a}^{(i)}_t}+(1-a^{(i)}_t)\log{(1-\hat{a}^{(i)}_t)})
\end{equation}
\end{small}

\subsubsection{Outcome prediction}
We finally adopt a potential outcome prediction network to estimate the the outcome $\hat{y}^{(i)}_{t,T+\tau}$ by taking the hidden representations $q^{(i)}_{t}$ from each time stamp as input. Here, to fully utilize the time series information, we use a max-pooling layer to aggregate all hidden representations. Let $g(\cdot)$ denotes the function learned from outcome prediction network. Then we have,
\begin{equation}
\label{eq:outcome-pred}
    \hat{y}^{(i)}_{t,T+\tau}=g(q^{(i)}_{m}, a^{(i)}=t)
\end{equation}
where we estimate the potential outcome for each patient given each treatment assignment situation. Here, we use MLPs to model the function $g(\cdot)$. We minimize the factual loss function as follows,
\begin{equation}
\label{eq:outcome-pred-loss}
    \mathcal{L}_y=\frac{1}{N}\sum_{i=1}^{N}w^{(i)}_a(\hat{y}^{(i)}_{t, T+\tau}-y^{(i)}_{t, T+\tau})^{2}
\end{equation}
where $w^{(i)}_a$ is to re-weight the population for adjusting confounders. We introduce the computation of $w^{(i)}_a$ in the following section.

\subsection{Balancing Module}
The predicted probability of receiving treatment is further leveraged for generating weights for each individual to balance the confounding. We compute the weights using inverse
probability of treatment weighting (IPTW) and extend to dynamic setting as follows,
\begin{equation}
\label{eq:weighting}
w^{(i)}_t=\frac{\text{Pr}(A)}{\hat{a}^{(i)}_t}+\frac{(1-\text{Pr}(A))}{(1-\hat{a}^{(i)}_t)}  
\end{equation}
where $\text{Pr}(A)$ denotes the probability of being in treated group and $\hat{a}^{(i)}_t$ is the predicted probability of receiving treatment given the current observed data and historical information. We take average of weights computed at each time stamp denoted as $w^{(i)}_a$.

\subsection{Loss Function}
The total loss function for the proposed method is defined as,
\begin{equation}
    \mathcal L=\mathcal{L}_y + \gamma \mathcal{L}_a + \lambda \lVert W \rVert_{2}
\end{equation}
where $\mathcal{L}_y$ is the factual prediction loss between estimated and observed factual outcomes, $\mathcal{L}_a$ is the loss from treatment prediction, $\gamma$, $\lambda$ are parameters to balance the loss function. The last term is $L_2$ regularization on model parameters $W$. The training process of \model is presented in Algorithm \ref{algorithm}.

\begin{algorithm}
	\caption{\model Model}
	\label{algorithm}
	\textbf{Input}: data of $i$-th patient: time-varying covariates $x^{(i)}_t$, static covariates $c$, treatment assignments $a^{(i)}_t$;\\
	\textbf{Output}: potential outcome $y^{(i)}_{T+\tau}$;
	\begin{algorithmic}[1]
    	\State Randomly initialize embedding matrix $W_{emb}$ for time-varying covariates, GRU parameters $W_f$, $W_r$, $W_h$, $V_f$, $V_r$, $b_f$, $b_r$, attention parameters $W_{\alpha}$;
    	\Repeat
    	\For{covariate v in $x^{(i)}_t$}
    	\State Obtain the embedding of v using Eq. (\ref{eq:embedding});
    	\State Obtain $q^{(i)}_t$ using Eq. (\ref{eq:hidden-representation-confounding});
    	\State Input the $q^{(i)}_t$ and $a^{(i)}_{t}$ into GRU and obtain $h^{(i)}_{t}$ using Eq. (\ref{eq:gru});
    	\State Compute the attention weights using Eq. (\ref{eq:attention}); (\ref{eq:context-vector})\;
    	\State Obtain $z^{(i)}_{t}$ using Eq. (\ref{eq:hidden-confounding})\;
    	\EndFor
    	\State Predict the treatment $\hat{a}^{(i)}_{t}$ using Eq. (\ref{eq:pred_a});
    	\State Compute the weights for each patient using Eq. (\ref{eq:weighting})
    	\State Calculate the treatment prediction loss using Eq. (\ref{eq:treatment-pred-loss});
    	\State Predict the potential outcome $\hat{y}^{(i)}_{T+\tau}$ using Eq. (\ref{eq:outcome-pred});
    	\State Calculate the outcome prediction loss using Eq. (\ref{eq:outcome-pred-loss});
    	\State Update parameters according to gradient of mean loss;
    	\Until{convergence}
	\end{algorithmic} 
\end{algorithm}

\vspace{-5pt}
\section{Experimental Setup}
To evaluate the performance of the proposed model, we conduct comprehensive comparison experiments on three different datasets: fully-synthetic dataset, semi-synthetic dataset and real-world dataset.

\subsection{Datasets and Simulation}
\subsubsection{Synthetic Dataset}
As introduced in the previous section, the treatment assignments $a^{(i)}_t$ at each time stamp are influenced by the confounders $q^{(i)}_t$, which are consist of previous hidden confounders $z^{(i)}_{t-1}$, current time-varying covariates $x^{(i)}_t$ and static features $c^{(i)}$. We first simulate $x^{(i)}_t$ and $z^{(i)}_{t}$ for each patient at time $t$ following $p$-order autoregressive process \cite{mills1991time} as,
\begin{equation}
\begin{aligned}
    x^{(i)}_{t,j} = \frac{1}{p}\sum_{r=1}^{p}(\alpha_{r,j}x^{(i)}_{t-r,j}+\beta_{r}a^{(i)}_{t-r}) + \eta_{t}\\
    z^{(i)}_{t,j} = \frac{1}{p}\sum_{r=1}^{p}(\mu_{r,j}z^{(i)}_{t-r,j}+\upsilon_{r}a^{(i)}_{t-r}) + \epsilon_{t}
\end{aligned}
\end{equation}
where $x^{(i)}_{t,j}$ and $z^{(i)}_{t,j}$ denote the $j$-th column of $x^{(i)}_t$ and $z^{(i)}_{t}$, respectively. For each $j$, $\alpha_{r,j},\mu_{r,j}\sim \mathcal{N}(1-(r/p),(1/p)^{2})$ control the amount of historical information of last p time stamps incorporated to the current representations. $\beta_{r},\upsilon_{r}\sim \mathcal{N}(0, 0.02^{2})$ controls the influence of previous treatment assignments. $\eta_{t},\epsilon_{t}\sim \mathcal{N}(0,0.01^{2})$ are randomly sampled noises. 

To simulate the treatment assignments, we generate $1000$ treated samples and $3000$ control samples. For treated samples, we randomly pick the treatment initial point among all time stamps. The treatments starting from the initial point are all set to $1$. For the control samples, the treatments at each time stamp are all set to $0$. 

The confounders $q^{(i)}_t$ at time stamp $t$ and outcome $y^{(i)}_{T+\tau}$ can be simulated using the hidden confounders and current covariates as follows,
\begin{equation}
\label{eq:synthetic-outcome}
\begin{aligned}
& q^{(i)}_t = \gamma_{h}\frac{1}{t}\sum_{r=1}^{t}z^{(i)}_{r} + (1-\gamma_{h})g([x^{(i)}_t,c^{(i)}])\\
& y^{(i)}_{T+\tau} = w^{\top}q^{(i)}_{T}+b
\end{aligned}
\end{equation}
    
where $\gamma_{h}$ is the parameter to control the influence of hidden confounders, $w\sim\mathcal{U}(-1,1)$ and $b\sim\mathcal{N}(0,0.1)$. The function $g(\cdot)$ maps the concatenated feature vectors $[x^{(i)}_t,c^{(i)}]$ into the hidden space. In this paper, for each individual, we simulate 100 time-varying covariates, 5 static covariates with 10 time stamps in total. We modify the value of $\gamma_{h}\in\{0.1, 0.3,0.5,0.7\}$ and obtain four variants of the current dataset. 

\subsubsection{Semi-synthetic Dataset based on MIMIC-III}
With a similar simulation process, we construct a semi-synthetic dataset based on a real-world dataset: Medical Information Mart for Intensive Care version III (MIMIC-III) \cite{johnson2016mimic}. MIMIC-III has more than 61,000 ICU admissions from 2001 to 2012 with recorded patients’ demographics and temporal information, including vital signs, lab tests, and treatment decisions. We extracted 11,715 adult sepsis patients fulfilling the sepsis-3 criteria \cite{sepsis3} as our studied cohort from MIMIC-III since sepsis contributes to up to half of all hospital deaths and is associated with more than \$24 billion in annual costs in the United States \cite{liu2014hospital}.

Here, we obtain 27 time-varying covariates (vital signs: temperature, pulse rate, glucose, etc; lab tests: potassium, sodium, chloride, etc.) and 12 static demographics (i.e., age, gender, race, height, weight, etc.) as potential confounding variables. The full list of covariates is available at Github\footnote{\url{https://github.com/ruoqi-liu/DSW}}. We consider vasopressors as treatments since they are commonly used in septic patients. As the MIMIC-III dataset is real world observational data, then it is impossible to obtain the counterfactual outcomes for calculating the ground truth treatment effect. Therefore, we simulate the potential outcomes for each patient using the observed covariates and treatment assignments. The simulation process is similar to the way we generate fully-synthetic dataset, with the exception that we only need to synthesize the potential outcomes (using Eq. \ref{eq:synthetic-outcome}). By varying the values of $\gamma_h\in\{0.1, 0.3, 0.5, 0.7\}$, we have four variants of the current datasets.

\subsubsection{Real world Dataset: MIMIC-III}
To evaluate the performance of our model in a real-world application, we design a causal inference setting based on the MIMIC-III dataset. (1) \textbf{Treatment}. We consider two available treatment assignments: vasopressors (vaso) and mechanical ventilator (mv). For each treatment option, we separately evaluate its causal effect on the important outcome signals. (2) \textbf{Outcomes}. To evaluate the treatment effect of vasopressors, we use mean blood pressure (Meanbp) as target outcomes since vasopressors are highly related to Meanbp. For mechanical ventilator, we adopt oxygen saturation (SpO2) as outcome since ventilator is usually assigned to patients with the difficulty of breathing. (3) \textbf{Confounders}. We consider the same confounders as in synthetic dataset (27 time-varying covariates and 12 static demographics).

\subsection{Methods for comparison}
To evaluate the performance of the proposed framework in estimating the ITE, we conduct comparison experiments on the following state-of-the-art causal inference methods,


\begin{itemize}[leftmargin=*]
    \item \textbf{Linear Regression (LR)}. LR directly regard the treatment as an additional feature for potential outcome prediction. It ignores the confounders and selection bias in observational data.
    \item \textbf{Random Forest (RF)}. The training process is same as LR (using treatment as a feature). We vary the number of trees in range $\{50,60,..,150\}$ and select best parameter setting on validation set. 
    \item \textbf{K-Nearest Neighbor Matching (KNN)} \cite{crump2008nonparametric}. KNN is a matching based method that estimates the counterfactual outcomes of treated (control) group from K-nearest neighbors in control (treated) group. We attempts three different distance metrics \textit{euclidean}, \textit{minkowski} and \textit{mahalanobis}, and choose the the metric yields best performance on the validation dataset.  
    \item \textbf{Propensity Score Matching (PSM)} \cite{rosenbaum1983central}. PSM is also a matching based method but instead use propensity score to measure the distance among individuals. Commonly, logistic regression is adopted for propensity score estimation. For logistic regression, we try different solvers: \textit{liblinear}, \textit{lbfgs}, \textit{sag} and \textit{saga} on the training set and select the solver with best performance on the validation set.
    \item \textbf{Counterfactual Regression (CFR)} \cite{johansson2016learning,shalit2017estimating}. CFR is deep representation learning based method. CFR has four different variants according the selection of distribution balancing metrics: Maximum Mean Discrepancy (\textbf{CFR MMD}), Wasserstein (\textbf{CFR WASS}), \textbf{BNN}, \textbf{TARNet}.
    \item \textbf{Causal Forest (CF)} \cite{wager2018estimation}. CF is an extension of RF for estimating the ITE, which is designed for causal effect estimation.
    \item \textbf{Bayesian Additive Regression Trees (BART)} \cite{hill2011bayesian}. BART is a non-parametric Bayesian regression tree model, which takes the covariates and treatment as inputs and outputs the distribution of outcomes. 
\end{itemize}


\subsection{Performance Measurement}
To evaluate the estimated ITE, we adopt mean squared error (MSE) between the ground truth and estimated ITE as follows,
\begin{equation}
    \text{PEHE} = \frac{1}{N}\sum_{i=1}^{N}((y^{(i)}_{1}-y^{(i)}_{0})-(\hat{y}^{(i)}_{1}-\hat{y}^{(i)}_{0}))^{2},
\end{equation}
which is also known as Precision in Estimation of Heterogeneous Effect (PEHE). Typically, we report the rooted PEHE $\sqrt{\text{PEHE}}$ in our paper. We are also interested in the causal effect over the whole population to help determine whether a treatment should be assigned to population. Then we calculate the mean absolute error (MAE) between the ground truth and estimated and average treatment effect (ATE):
\begin{equation}
    \text{ATE}=|\frac{1}{N}\sum_{i=1}^{N}(y^{(i)}_{1}-y^{(i)}_{0})-\frac{1}{N}\sum_{i=1}^{N}(\hat{y}^{(i)}_{1}-\hat{y}^{(i)}_{0})|
\end{equation}
Additionally, we adopt rooted mean squared error (RMSE) between the estimated factual outcomes and ground truth outcomes to evaluate the performance on factual prediction task as follows,
\begin{equation}
\text{RMSE} = \sqrt{\frac{1}{N}\sum_{i=1}^{N}(\hat{y}^{(i)}_{t}-y^{(i)}_{t})^2}
\end{equation}


\subsection{Implement Details}
The model is implemented and trained with Python 3.6 and PyTorch 1.4 \footnote{https://pytorch.org/}, on a high-performance computing cluster with four NVIDIA TITAN RTX 6000 GPUs. We train our model using the adaptive moment estimation (Adam) algorithm with a batch size of 128 subjects and the learning rate is 0.001. The data is randomly split into training, validation and test sets with a ratio of $70\%$, $10\%$, $20\%$. The information from a given patient is only present in one set. The validation set is used to improve the models and select the best model hyper-parameters. We report the performance of our model and baselines on the test set. We use $\sqrt{\text{PEHE}}$ and ATE error to measure the models' performance on ITE estimation, and RMSE for factual prediction. As all baseline methods are originally designed for static environment, we run these models independently on each time stamp and average the evaluation metrics over all time stamps. The code and more implementation details are available at \url{https://github.com/ruoqi-liu/DSW}.

\section{Results}
We now report the performance of \model on synthetic, semi-synthetic and real-world datasets. We focus on answering the following research questions by our experimental results:
\begin{itemize}[leftmargin=*]
    \item \textbf{Q1: How precise is \model on ITE estimation?}
    \item \textbf{Q2: How accurate is \model on factual prediction task (i.e., outcome prediction)?}
    \item \textbf{Q3: How can \model be used for personalized medicine?}
\end{itemize}

\subsection{How precise is \model on ITE estimation?}
We conduct comprehensive comparison experiments on synthetic and  semi-synthetic datasets and report $\sqrt{\text{PEHE}}$ and ATE on each dataset. By varying the parameter $\gamma_{h}$ that controls the influence of hidden confounders, we evaluate the how precise is \model on ITE estimation under different value of $\gamma_{h}$. 

\noindent\textbf{Results on Synthetic Dataset}
Table \ref{tab:main_results_syn} shows the performance of our method and baselines on fully-synthetic dataset evaluated by $\sqrt{\text{PEHE}}$ and ATE. The values shown in the table are averaged on 10 realizations. We observe that \model outperforms all other baselines with different value of $\gamma_{h}$, which confirms that our designed framework can better capture the characteristics of longitudinal data and generate accurate estimation of ITE. 

Generally speaking, the representation learning based approaches achieve better performance compared with base methods, matching based methods and tree based methods. Since those linear approaches are not designed for causal effect estimation, they may not able to control the influence of confounding variables. The matching based methods consider the similarity information among treated and control groups to alleviate the selection biases. However, their estimation becomes inaccurate when dealing with high-dimensional and complex data. Tree and forest based method achieve comparable performance with basic random forest method since these two methods are established upon the random forest.

The representation learning based methods use the deep neural network to model the representations of confounding variables. Their methods achieve the best performance among all baselines. CFR MMD, CFR WASS, BNN and TARNet share the similar design of neural network, with exception that they adopt different strategies to minimize the distance between treated and control groups. Specifically, CFR MMD and CFR WASS have the same outcome prediction networks, but the former uses Maximum Mean Discrepancy (MMD) and the latter uses Wasserstein (WASS) to balance the distributions. BNN regards the treatment as additional feature and minimize the distances between treated and control group in latent space. TARNet is a vanilla version without balancing property. Among all these four representation learning based methods, we observe that CFR MMD and CFR WASS generally achieve better performance than other two methods. Although representation learning based models outperform the other baselines, they ignore the time-varying confounders and lose lots of temporal information, which are crucial and common in healthcare data. Our proposed \model successfully captures the temporal information and thus outperform the baselines.

Note that, four variants of simulated datasets ($\gamma_{h}\in\{0.1,0.3,0.5,0.7\}$) are not comparable, since the distribution of simulated outcomes are not same. With the increasing of $\gamma_h$, less portion of observed confounders are included, which results in smaller values of outcomes. Instead, we focus on the performance of methods within each dataset. 


\begin{table*}[t]
\centering
\caption{Performance comparison on synthetic datasets. We construct four variants of datasets by varying the value of $\gamma_h\in\{0.1,0.3,0.5,0.7\}$. Here, we report the estimated $\sqrt{\text{PEHE}}$ and ATE of each method among four datasets.}
\label{tab:main_results_syn}
{\renewcommand{\arraystretch}{1.2}
\begin{tabular}{lllllllllll}
\hline
 & &
  \multicolumn{2}{c}{$\gamma_{h}=0.1$} &
  \multicolumn{2}{c}{$\gamma_{h}=0.3$} &
  \multicolumn{2}{c}{$\gamma_{h}=0.5$} &
  \multicolumn{2}{c}{$\gamma_{h}=0.7$} \\ \cline{3-10} 
& Method &
  \multicolumn{1}{c}{$\sqrt{\text{PEHE}}$} &
  \multicolumn{1}{c}{ATE} &
  \multicolumn{1}{c}{$\sqrt{\text{PEHE}}$} &
  \multicolumn{1}{c}{ATE} &
  \multicolumn{1}{c}{$\sqrt{\text{PEHE}}$} &
  \multicolumn{1}{c}{ATE} &
  \multicolumn{1}{c}{$\sqrt{\text{PEHE}}$} &
  \multicolumn{1}{c}{ATE} \\ \hline
\multirow{2}{*}{Base model} &    LR            & 0.640 & 0.551 & 0.648 & 0.558 & 0.656 & 0.566 & 0.663 & 0.574 \\
& Random Forest & 0.656 & 0.552 & 0.658 & 0.553 & 0.660 & 0.555 & 0.663 & 0.557 \\ \hline
\multirow{2}{*}{Matching based} & KNN \cite{crump2008nonparametric}           & 0.713 & 0.604 & 0.718 & 0.608 & 0.724 & 0.611 & 0.729 & 0.615 \\
& PSM \cite{rosenbaum1983central}           & 0.699 & 0.591 & 0.708 & 0.602 & 0.714 & 0.607 & 0.720 & 0.611 \\ \hline
  & CFR MMD \cite{shalit2017estimating}       & 0.587 & 0.485 & 0.594 & 0.491 & 0.597 & 0.492 & 0.600 & 0.495 \\
Representation & CFR WASS \cite{shalit2017estimating}      & 0.571 & 0.470 & 0.574 & 0.473 & 0.576 & 0.474 & 0.580 & 0.476 \\
learning based & BNN \cite{johansson2016learning}          & 0.586 & 0.488 & 0.593 & 0.495 & 0.595 & 0.496 & 0.597 & 0.498 \\
& TARNet \cite{shalit2017estimating}       & 0.606 & 0.512 & 0.610 &0.516 & 0.615 & 0.520 & 0.619 & 0.523 \\ \hline
\multirow{2}{*}{Forest based}   
&  Causal Forest \cite{wager2018estimation} & 0.604 & 0.511 & 0.612 & 0.517 & 0.615 & 0.521 & 0.618 & 0.523 \\
& BART \cite{hill2011bayesian}   & 0.608 & 0.520 & 0.614 & 0.525 & 0.621 & 0.530 & 0.619 & 0.528 \\ \hline
Ours & \textbf{\model} &
  \textbf{0.491} & \textbf{0.391} & \textbf{0.469} & \textbf{0.372} & \textbf{0.485} & \textbf{0.384} & \textbf{0.515} & \textbf{0.397} \\ 
  \hline
\end{tabular}}
\vspace{-10pt}
\end{table*}

\noindent\textbf{Results on Semi-synthetic Dataset}
We demonstrate the performance of the proposed model and baselines on a semi-synthetic dataset based on MIMIC-III. As shown in Table \ref{tab:main_results_mimic_syn}, \model achieves the best performance among other baseline methods in terms of $\sqrt{\text{PEHE}}$ and ATE. We vary the value of $\gamma_h$ to control the influence of hidden confounders and conduct comparison experiments under each value of $\gamma_h$. 

We observe that representation learning based methods in general outperform most causal inference methods which demonstrate that deep learning architecture with distribution balancing is beneficial to the ITE estimation. Base models (LR and Random Forest) cannot perform very well since they ignore the selection bias and confounding factors. As for matching based methods, they consider the similarity information among individuals to alleviate selection bias. Among all the baselines, \model considers the temporal information in the time-varying data, and thus generate unbiased and accurate treatment effect estimation.

\begin{table*}[t]
\centering
\caption{Performance comparison on semi-synthetic MIMIC-III datasets. We construct four variants of datasets by varying the value of $\gamma_h\in\{0.1,0.3,0.5,0.7\}$. Here, we report the estimated $\sqrt{\text{PEHE}}$ and ATE of each method among four datasets.}
\label{tab:main_results_mimic_syn}
{\renewcommand{\arraystretch}{1.2}
\begin{tabular}{llllllllll}
\hline
&               & \multicolumn{2}{c}{$\gamma_{h}=0.1$} & \multicolumn{2}{c}{$\gamma_{h}=0.3$} & \multicolumn{2}{c}{$\gamma_{h}=0.5$} & \multicolumn{2}{c}{$\gamma_{h}=0.7$} \\ \cline{3-10} 
 &
  Method &
  \multicolumn{1}{c}{$\sqrt{\text{PEHE}}$} &
  \multicolumn{1}{c}{ATE} &
  \multicolumn{1}{c}{$\sqrt{\text{PEHE}}$} &
  \multicolumn{1}{c}{ATE} &
  \multicolumn{1}{c}{$\sqrt{\text{PEHE}}$} &
  \multicolumn{1}{c}{ATE} &
  \multicolumn{1}{c}{$\sqrt{\text{PEHE}}$} &
  \multicolumn{1}{c}{ATE} \\ \hline
\multirow{2}{*}{Base model}     & LR            & 0.823    & 0.660     & 1.414      & 1.134    & 1.474     & 1.233    & 1.542     & 1.336     \\
& Random Forest & 0.837    & 0.671     & 1.422      & 1.143    & 1.484     & 1.245     & 1.553     & 1.349     \\ \hline
\multirow{2}{*}{Matching based} & KNN \cite{crump2008nonparametric}           & 0.818    & 0.650     & 1.408      & 1.137    & 1.472     & 1.237     & 1.543     & 1.337     \\
& PSM \cite{rosenbaum1983central}          & 0.767    & 0.607    & 1.417     & 1.143   & 1.477     & 1.240    & 1.548     & 1.340     \\ \hline
\multirow{4}{*}{\begin{tabular}[c]{@{}l@{}}
Representation \\ learning based\end{tabular}} &
  CFR MMD \cite{shalit2017estimating} & 0.803  &
  0.643 & 1.257 & 0.972 &  1.356 & 1.099 &
  1.527 & 1.318 \\
& CFR WASS \cite{shalit2017estimating}    & 0.800    & 0.641    & 1.256      & 0.972    & 1.370     & 1.118     & 1.556     & 1.352     \\
& BNN \cite{johansson2016learning}           & 0.802    & 0.643     & 1.287      & 1.006    & 1.337     & 1.080    & 1.515     & 1.305     \\
& TARNet \cite{shalit2017estimating}        & 0.829    & 0.664     & 1.256      & 0.972    & 1.373     & 1.120     & 1.527     & 1.318     \\ \hline
\multirow{2}{*}{Forest based}   
& Causal Forest \cite{wager2018estimation} & 0.796    & 0.639     & 1.413      & 1.134    & 1.473     & 1.233    & 1.540     & 1.335     \\
& BART \cite{hill2011bayesian}        & 0.785    & 0.631    & 1.413     & 1.135    & 1.472   & 1.234     & 1.538     & 1.336     \\ \hline
Ours &
  \textbf{\model} &
  \textbf{0.522} &
  \textbf{0.432} &
  \textbf{0.604} &
  \textbf{0.523} &
  \textbf{0.672} &
  \textbf{0.601} &
  \textbf{0.722} &
  \textbf{0.652} \\ \hline
\end{tabular}}
\vspace{-10pt}
\end{table*}

\begin{table}[t]
\centering
\caption{Factual Prediction on the MIMIC-III dataset. We select two treatment-outcome pairs:  Vasopressor-Meanbp and Ventilator-SpO2, and report RMSE between estimated factual outcome and ground truth. }
\label{tab:factual_prediction}
{\renewcommand{\arraystretch}{1.3}
\begin{tabular}{lcc}
\hline
              & \multicolumn{2}{c}{MIMIC Dataset (RMSE)} \\ \cline{2-3} 
Method        &Vent-SpO2           & Vaso-Meanbp         \\ \hline
LR & 0.909 &0.973             \\
KNN \cite{crump2008nonparametric}  & 0.901  & 1.030             \\
CFR WASS \cite{shalit2017estimating}  & 0.870  & 1.011          \\
BART \cite{hill2011bayesian}  & 0.873  & 0.966           \\ \hline
\textbf{\modelnospace} & \textbf{0.814} & \textbf{0.814} \\ \hline
\end{tabular}}
\vspace{-20pt}
\end{table}


\vspace{-3pt}
\subsection{How accurate is \model on factual prediction?}
In real-world data, we have no access to counterfactual outcomes for calculating the true treatment effect, and thus we cannot compute the $\sqrt{\text{PEHE}}$ and ATE. Instead, we evaluate the performance of our model through a factual prediction task. We measure the RMSE between observed (factual) outcomes and estimated outcomes on two treatment-outcome pairs: vasopressor-meanBP and ventilator-SpO2.


Table \ref{tab:factual_prediction} shows the estimated RMSE of two selected pairs. Since causal effect methods are not initially designed for factual prediction, we adopt four baselines from each category that can be adapted for factual prediction task: LR, KNN, CFR (WASS) and BART. Matching based methods aim to estimate the counterfactual and cannot be directly used for factual inference. We adapt the KNN matching for factual prediction by combining the treatments as additional features to predict factual outcomes. Among four representation learning based methods, they achieve relative comparative performance in two datasets, so we use CFR WASS as a representative for this kind of method. In two forest based methods, CF directly outputs the estimated ITE without any inference of factual outcomes, so we use BART as our baseline. 


As shown in TABLE \ref{tab:factual_prediction}, \model outperforms all the baselines, which demonstrates that our model yields accurate estimation on factual data. Among the baselines, we find that representation learning based method (CFR WASS) performs better than the linear regression method (LR) and matching based method (KNN), which is consistent with TABLE \ref{tab:main_results_syn} and \ref{tab:main_results_mimic_syn}. 


\subsection{How can \model be used for personalized medicine?}
Based on observational data, we are going to show that our model can adjust time-varying confounders, and generate unbiased and accurate estimation of treatment effect.
In this case, our model could potentially help physicians determine whether to apply a specific treatment to a patient. We examine the usages of \textit{Vasopressor} and \textit{Ventilator} in the analysis, which are commonly used treatments for septic patients.

\begin{figure*}[t]
\centering
\subfigure[]{\label{fig:vaso-below}
\includegraphics[width=0.34\textwidth]{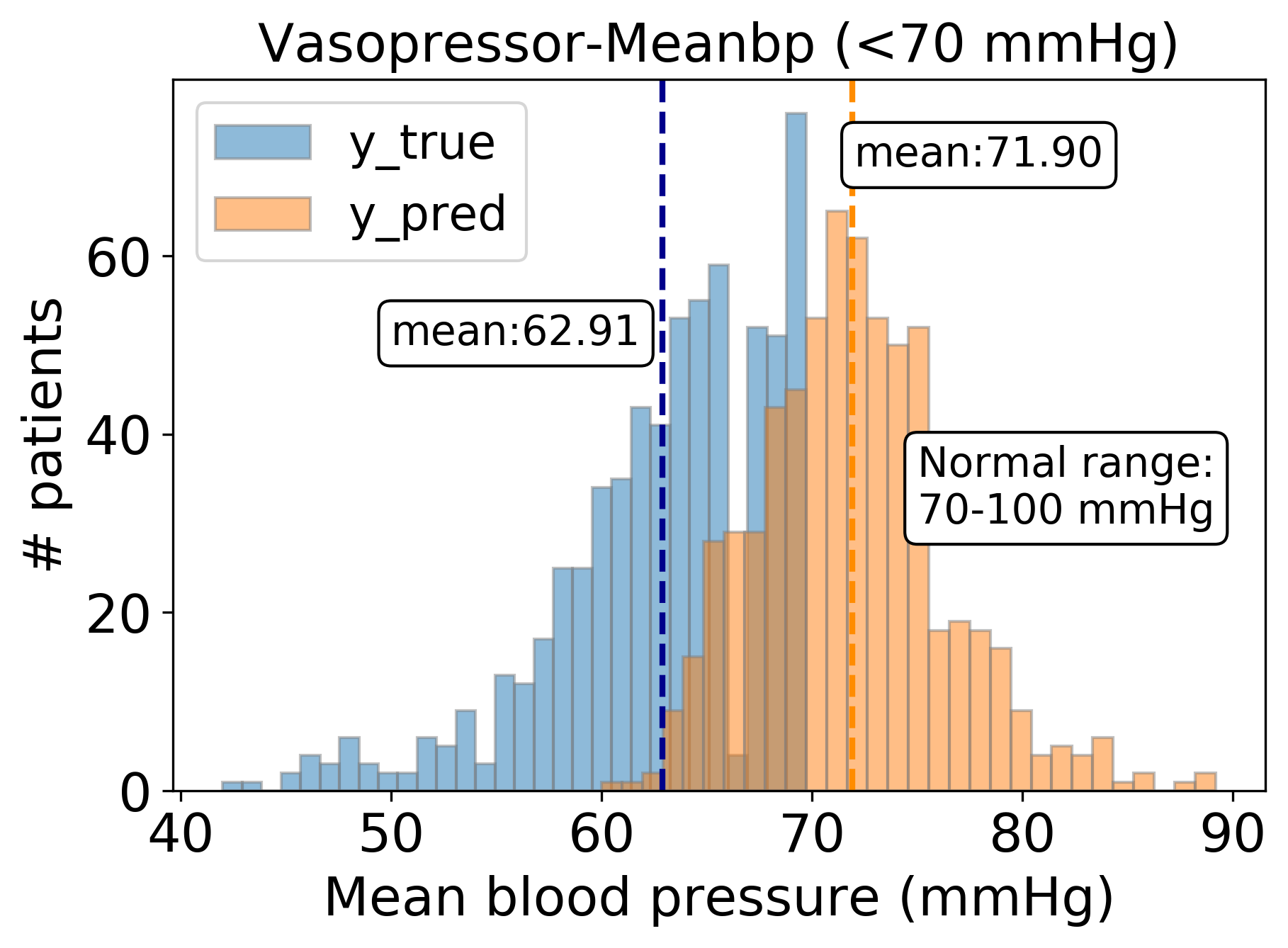}}
\subfigure[]{\label{fig:vaso-above}
\includegraphics[width=0.34\textwidth]{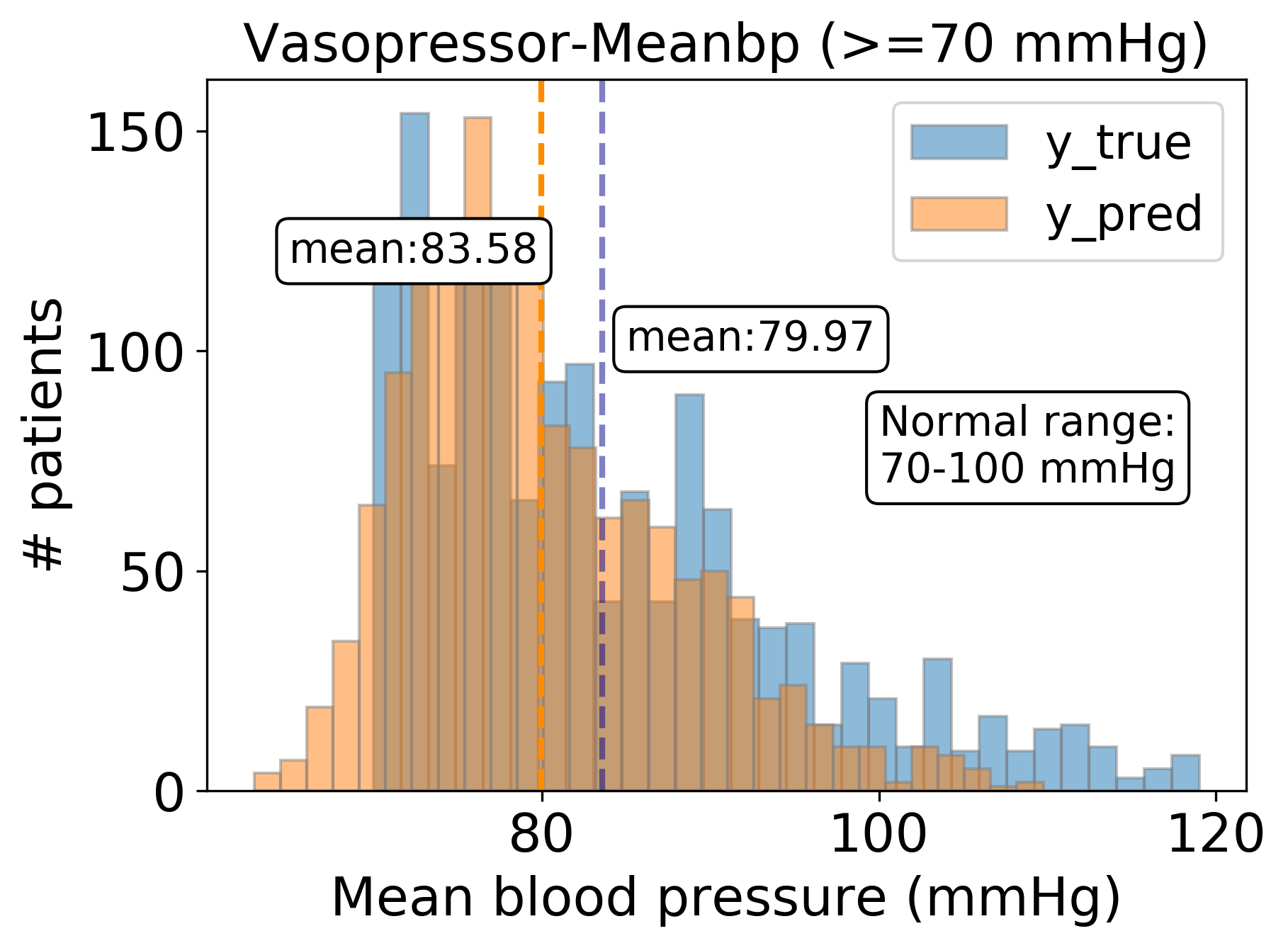}}
\vspace{-4pt}
\subfigure[]{\label{fig:vent-below}
\includegraphics[width=0.34\textwidth]{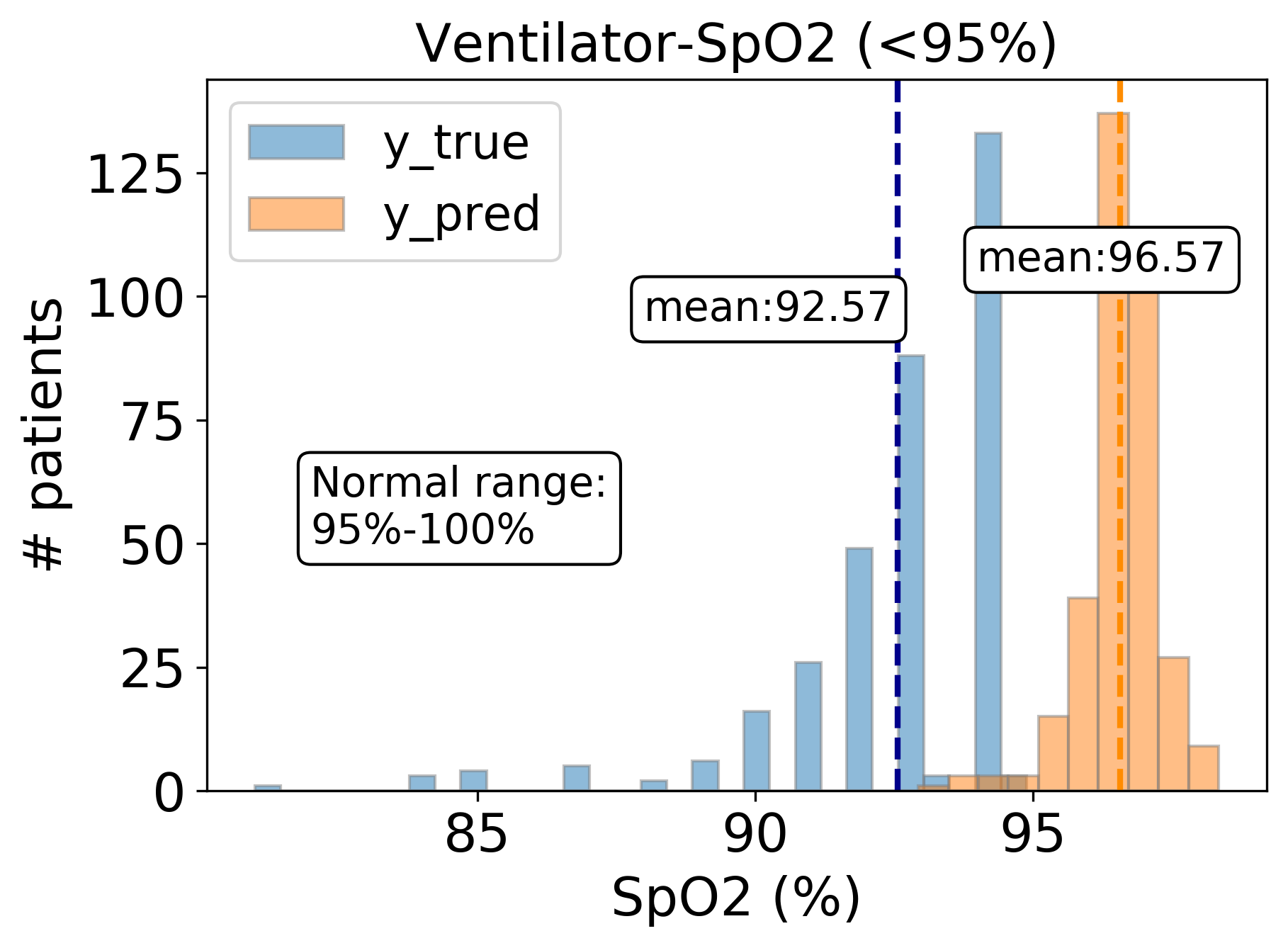}}
\subfigure[]{\label{fig:vent-above}
\includegraphics[width=0.34\textwidth]{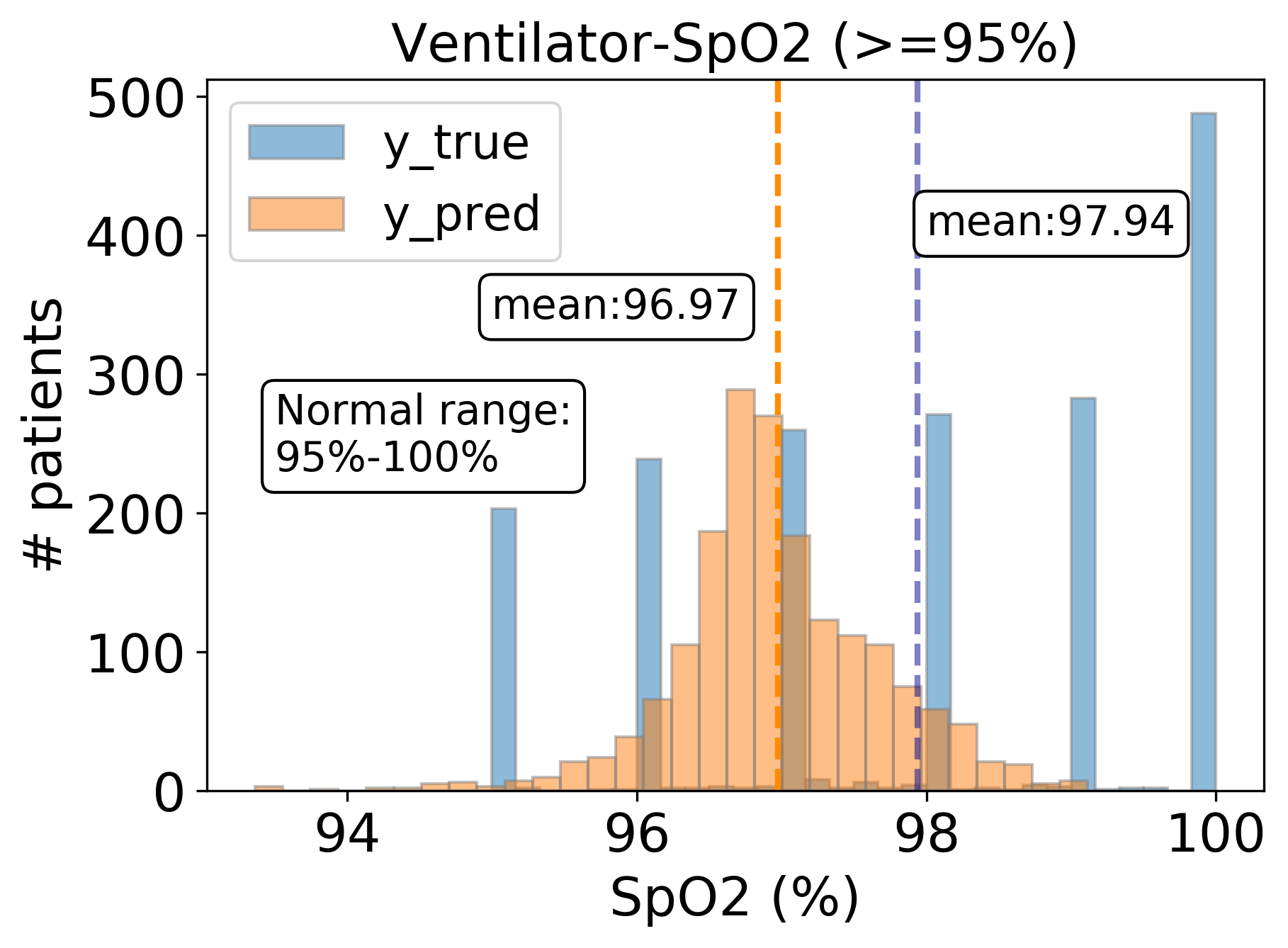}}
\caption{Distribution of ground truth MeanBP and predicted MeanBP given vasopressor as treatment. Fig. \ref{fig:vaso-below} displays the distribution of patients with observed MeanBP below than 70 mmHg. Fig. \ref{fig:vaso-above} displays the distribution of patients with observed MeanBP above 70 mmHg. Fig. \ref{fig:vent-below} displays the distribution of patients with observed SpO2 below than 95\%. Fig. \ref{fig:vent-above} displays the distribution of patients with observed SpO2 above 95\%.}
\vspace{-10pt}
\label{fig:case-study}
\end{figure*}


\vspace{3pt}
\noindent\textbf{Vasopressor-Meanbp pair}
Vasopressor (a.k.a., antihypotensive agent) is a group of medications that tend to raise low blood pressure. The patients are expected to have normal blood pressure after receiving a vasopressor. To demonstrate that our model can adjust time-varying confounders, we plot the distribution of ground truth (observed) and predicted Meanbp values after the patients have received vasopressor in Fig. \ref{fig:case-study}. According to the threshold of the normal range of Meanbp (70-100 mmHg), we separate the patients into two groups: one is the patients with observed MeanBP values below 70 mmHg, the other is the patients with observed Meanbp values above 70 mmHg. Figure. \ref{fig:vaso-below} shows the distribution of patients with Meanbp below 70 mmHg. We observe this group of patients remains low blood pressure even after receiving vasopressor and the average value is far lower than 70 mmHg. If mainly based on the observed data, we may conclude that vasopressor has no effect on raising blood pressure and are unnecessary to be assigned to patients with low blood pressure. However, the predicted values given by our model are higher than the observed values and the average Meanbp belongs to the normal range, which indicates that vasopressor should have a beneficial effect on the blood pressure. Figure \ref{fig:vaso-above} shows the distribution of patients with Meanbp above 70 mmHg. For these patients whose Meanbp remains in the normal range after receiving vasopressor, the predicted values are still within the normal range, which indicates that vasopressor should be assigned to this group of patients to maintain normal blood pressure. If not, their situation may become worse.    

\vspace{3pt}
\noindent\textbf{Ventilator-SpO2 pair}
A ventilator is a machine that delivers breaths to a patient who is physically unable to breathe, or breathing insufficiently to maintain blood oxygen. We monitor the value of oxygen saturation (SpO2) to estimate the treatment effect. 

Similarly, we show the distribution of ground truth and predicted SpO2 values of patients who have received a ventilator during the observational window in Fig. \ref{fig:case-study}. As the normal range of SpO2 is 95-100\%, we separately plot the distribution of patients with observed SpO2 values below 95\% in Fig. \ref{fig:vent-below}, and the distribution of patients with observed SpO2 values above 95\% in Fig. \ref{fig:vaso-above}. We observe that, for patients with observed SpO2 lower than normal value, the distribution of predicted SpO2 values lies in normal range with an average of 96.57\%. And for patients with observed SpO2 in the normal range, our predicted values still belong to the normal range. 

Results show that our model adjusts time-varying confounders and is able to generate unbiased and accurate ITE on important outcome signals (vasopressor's effect on blood pressure and ventilator's effect on blood oxygen in our analysis). Thus, it could potentially assist physicians to determine whether to introduce a treatment to a specific patient, paving the way for personalized medicine.

\section{Related Work}
In this section, we review the related work for ITE estimation using static and time-varying observational data. We first introduce the causal effect learning framework on static data, and then the framework based on time-varying data. 

\noindent\textbf{Learning causal effects with static data} According to the way to control the confounders, existing work with static observational data can be divided into four groups: 1) Matching-based methods; 2) Tree-based methods; 3) Reweighting-based methods; 4) Representation-based methods. The matching-based methods are adopted to estimate the counterfactual from the nearest neighbors. The distances among individuals can be measured in several ways (i.e., Euclidean distance, propensity scores). For example, propensity score matching (PSM) \cite{rosenbaum1983central} is to match a treated (control) sample to a set of control (treated) samples with similar propensity scores. Tree-based methods are also widely adopted in causal effect estimation. Bayesian additive regression trees (BART) \cite{hill2011bayesian} is a non-parametric Bayesian regression tree model based on the \textit{strong ignorability assumption}. It is easy to implement and free from parameter adjustment. Causal Forest (CF) \cite{wager2018estimation} is also a tree-based causal effect estimation method, which estimates the treatment effect at the leaf node by mapping the original covariate into tree and forests. The reweighting-based methods attempt to re-weight samples in the population for correcting the bias in observational data. For example, inverse probability of treatment weighting (IPTW) \cite{rosenbaum1983central} removes the confounding by assigning a weight to each individual in the population. The weights are calculated based on the propensity score. Recently, representation learning methods are proposed for causal effect estimation via balancing the distribution between treated and control groups in hidden space \cite{johansson2016learning,shalit2017estimating}. Moreover, Yao et al. \cite{yao2018representation} incorporate the local similarity among individuals with population-level distribution balancing in latent space to better estimate ITE. Yoon et al. propose to use generative adversarial nets (GAN) for inferring the counterfactual outcomes based on factual outcomes. Shi et al. \cite{shi2019adapting} jointly model the propensity prediction and potential outcome prediction as a multi-task learning problem. 

Though existing work shows great performance in causal effect estimation, they still have some limitations. First, most of them are built upon \textit{strong ignorability assumption} without considering the influence of hidden confounders. This constrain has been shown to lead to bias in estimating causal effects \cite{pearl2009causality}. Moreover, existing work is initially designed for static data, which is not easy to adapt for ITE estimation under dynamic longitudinal setting. 

\vspace{3pt}
\noindent\textbf{Learning causal effects with time-varying data}
As estimating causal effect from observational data is significant and most observational data contains sequential information, some work has been proposed for dealing time-varying confounders. In statistics and epidemiology domains, a group of methods use the inverse probability of treatment and \textit{g-formula} based method to estimate causal effect with sequential data \cite{robins2000marginal,schulam2017reliable}. More recently, Lim et al. \cite{lim2018forecasting} propose a recurrent marginal structural network for predicting the patient's potential response to a series of treatments. Bica et al. \cite{bica2020estimating} adopt adversarial training techniques to balance the historical confounding variables. Their method is based on the \textit{strong ignorability assumption}. Later, Bica et al. \cite{bica2019time} relax the assumption on strong ignorability and propose to estimate the treatment response with the existence of hidden confounders. 
\section{Conclusion}
In this paper, we propose Deep Sequential Weighting (\modelnospace), a deep learning based framework for estimating ITE with time-varying confounders. Specifically, \model infers the hidden confounders by incorporating the current treatment assignments and historical information using a deep recurrent weighting neural network. When combined with current observed data, the learned representations of hidden confounders are leveraged for potential outcome prediction and treatment prediction. We compute the time-varying inverse probabilities of treatment for re-weighting the population. Comprehensive experiments on fully-synthetic, semi-synthetic, and real-world datasets demonstrate the effectiveness of \model when compared to state-of-the-art baseline methods. Results illustrate that our model can generate unbiased and accurate treatment effect by conditioning on time-varying confounders. Our model has the potential to be used as part of clinical decision support systems to determine whether a treatment is needed for a specific patient, paving the way for personalized medicine.

\end{document}